\documentclass[aps,pra,notitlepage]{revtex4-1}
\usepackage[english]{babel}
\usepackage{amsmath,amssymb,graphicx,bbm}

\usepackage{times}

\newcommand{\nobracket}{}
\newcommand{\nocomma}{}
\newcommand{\noplus}{}
\newcommand{\tmem}[1]{{\em #1\/}}
\newcommand{\tmmathbf}[1]{\ensuremath{\boldsymbol{#1}}}
\newcommand{\tmop}[1]{\ensuremath{\operatorname{#1}}}
\newcommand{\tmtextbf}[1]{{\bfseries{#1}}}

\begin{document}

\title{Reduced dynamical maps in the presence of initial correlations}

\author{Bassano Vacchini}
\affiliation{Dipartimento di Fisica, Universit\`a degli Studi di Milano,
Via Celoria 16, I-20133 Milan, Italy}
\affiliation{INFN, Sezione di Milano, Via Celoria 16, I-20133 Milan, Italy}

\author{Giulio Amato}
\affiliation{Dipartimento di Fisica, Universit\`a degli Studi di Milano,
Via Celoria 16, I-20133 Milan, Italy}

\begin{abstract}
  We introduce a general framework for the construction of completely positive
  dynamical evolutions in the presence of system-environment initial
  correlations. The construction relies upon commutativity of the
  compatibility domain obtained by considering the marginals with respect to
  the environmental degrees of freedom of the considered class of correlated
  states. Our approach allows to consider states whose discord is not
  necessarily zero and explicitly show the non-uniqueness of the completely
  positive extensions of the obtained dynamical map outside the compatibility
  domain. The relevance of such maps for the treatment of open quantum system
  dynamics is discussed and connection to previous literature is critically
  assessed.
\end{abstract}

{\maketitle}

\section{Introduction}

A ubiquitous situation in quantum physics involves the description of systems
which are not isolated, so that their dynamics is actually influenced by other
quantum degrees of freedom. The theory of open quantum systems was indeed
developed to cope with such situations and has found important applications in
diverse fields starting from quantum optics to condensed matter theory,
chemical physics and many others {\cite{Breuer2002,Weiss1999a}}. The standard
description of an open quantum system dynamics rests on two basic assumptions,
namely an initial system-environment state in factorized form and weak
coupling between system and environment. In such a case the existence of a
reduced dynamics is granted and it can reasonably be taken to obey a semigroup
composition law in time, so that the most famous result by Gorini,
Kossakowski, Sudarshan and Lindblad applies, fully characterizing the
generator of such a dynamical semigroup {\cite{Gorini1976a,Lindblad1976a}}. A
lot of effort has been devoted to overcome these limitations within the
standard framework in which the reduced open quantum system dynamics is
obtained by tracing over the environmental degrees of freedom, but also
different approaches have been recently proposed {\cite{Bondar2014a}}.

Major results have been obtained in describing dynamics beyond the weak
coupling limit, leading to reduced dynamical evolutions which go beyond a
simple semigroup composition law. These strong coupling dynamics often show up
memory effects. Indeed also in this respect important results have been
obtained in providing a characterization of non-Markovian dynamics within open
quantum system theory, and all of these results rely on the existence of a
reduced system dynamics {\cite{Rivas2014a,Breuer2016a}}. On the contrary,
despite important efforts
{\cite{Rodriguez2008a,Shabani2009a,Rodriguez2010a,Brodutch2013a,Sabapathy2013a,Liu2014a,Dominy2016a}},
the extension of the formalism to go beyond initially factorized states
appears to be much harder, and a general satisfactory treatment still lags
behind. Indeed such an extension is called for, since the choice of factorized
initial states is well compatible with a weak coupling approach, but is
generally not a natural assumption if one considers situations in which the
coupling between system and environment degrees of freedom is actually strong.

Different paths have been followed in order to tackle the issue of initially
correlated states between system and environment
{\cite{Grabert1988a,Devi2011a,Modi2012a,Ignatyuk2013a,Breuer2006a}}, and in
particular great attention has been devoted to study the conditions under
which a completely positive map can be introduced to describe the reduced
dynamics in the presence of initial correlations
{\cite{Pechukas1994a,Alicki1995a,Shabani2009a,Stelmachovic2001a,Rodriguez2010a,Buscemi2014a}}.

In the usual framework of open quantum systems {\cite{Breuer2002}} one
considers a tensor product structure $\mathcal{H}_S \otimes \mathcal{H}_E
\nocomma$, where $\mathcal{H}_S$ and $\mathcal{H}_E$ denote the Hilbert space
of system and environment respectively, and assumes that the overall system is
closed, so that its time evolution can be described by a group of unitary
operators $U_{SE} (t)$. Within this description, given an arbitrary initial
system-environment state $\rho_{SE} (0)$, with corresponding reduced system
state $\rho_S (0) = \tmop{Tr}_E \rho_{SE} (0)$, where $\tmop{Tr}_E$ denotes
the partial trace with respect to the environment degrees of freedom, one can
naturally consider the following collection of time dependent transformations
\begin{eqnarray}
  \rho_S (0) & \rightarrow & \rho_S (t) = \tmop{Tr}_E [U_{SE} (t) \rho_{SE}
  (0) U_{SE} (t)^{\dag}], \label{eq:map} 
\end{eqnarray}
which by construction preserve positivity and trace. If the initial state
$\rho_{SE} (0)$ is actually factorized, so that $\rho_{SE} (0) = \rho_S (0)
\otimes \rho_E (0)$, it is well known that in such a way one obtains a linear
map defined on the whole set of states, which in particular can be shown to be
not only positive, but actually completely positive. The notion of complete
positivity {\cite{Nielsen2000,Holevo2001}} naturally emerges in this open
quantum system setting and is indeed a typical quantum feature, related to the
tensor product structure of the space describing a composite system. At the
level of states one can witness the difference between positivity and complete
positivity of a map by the application to entangled states, while at the level
of observables the same difference can be appreciated by applying the map to
non commuting set of observables. Indeed it is an important result that for a
map acting on a commutative space positivity is equivalent to complete
positivity {\cite{Stinespring1955a,Takesaki2002}}. More precisely the notions
of positivity and complete positivity coincide if either starting or arrival
space of the map is given by a commutative algebra, which therefore is
amenable to a classical description {\cite{Strocchi2005}}.

In this article we will build on these basic facts to point out a general
construction of completely positive maps arising in the presence of a
correlated system environment state. As we shall see this approach allows us
to recover as special cases some results previously obtained in the literature
{\cite{Rodriguez2008a,Brodutch2013a}}.

\section{Quantum maps and correlated
states}

\subsection{General construction of quantum maps starting from correlated
states}

In order to consider the possibility to introduce completely positive maps
starting from correlated system environment states let us first consider the
following class of correlated states
\begin{eqnarray}
  \mathcal{C}_{SE} & = & \left\{ \rho_{SE} = \sum_i p_i \sigma^i_S \otimes
  \rho^i_E \quad s.t. \quad [\sigma^i_S, \sigma^j_S] = 0 \right\},
  \label{eq:cse} 
\end{eqnarray}
where $\{ p_i \}_i$ is a probability distribution, the $\{ \rho^i_E \}_i$ a
collection of states for the environment, $\{ \sigma^i_S  \}_i$ a fixed set of
commuting statistical operators for the system and we assume the Hilbert space
of the system $\mathcal{H}_S$ to be finite dimensional with dimension $n$. The
set $\mathcal{C}_{SE}$ provides a convex subset of the whole set of states on
$\mathcal{H}_S \otimes \mathcal{H}_E$, which we denote by $\mathcal{S}
(\mathcal{H}_S \otimes \mathcal{H}_E)$. In particular it is a subset of the
set of separable states which includes only zero discord states
{\cite{Ollivier2001a,Henderson2001a}}. To this set we can associate a
compatibility domain given by the set of system states which can be obtained
as marginals of these correlated states, namely
\begin{eqnarray}
  \mathcal{C}_S & = & \left\{ \rho_S = \sum_i p_i \sigma^i_S \right\},
  \label{eq:cs} 
\end{eqnarray}
which is still a convex set and is in particular a commutative set. Note
however that the relationship between sets of correlated states and their
compatibility domain is many to one, so that the same compatibility domain may
arise from different classes of separable correlated states. The set \ $\{
\sigma^i_S  \}_i$ is generated by a set of $d \leqslant n$ statistical
operators with orthogonal support, where $d$ is the dimension of the linear
hull of $\mathcal{C}_S$. If $d = n$ in particular $\mathcal{C}_S$ is given by
the convex combinations of states which are all extremal. Supposing $d < n$,
so that at least one of the $\{ \sigma^i_S  \}_i$, say $W$, is not necessarily
a projection operator, without loss of generality we have
\begin{eqnarray}
  \{ \sigma^i_S  \}_i & = & \{ \Pi_{\varphi_1} \nocomma, \ldots,
  \Pi_{\varphi_{d - 1}}, W \}, \label{eq:ssi} 
\end{eqnarray}
where $\{ \varphi_i \}_{i = 1, \ldots, d - 1}$ are orthonormal vectors in
$\mathcal{H}_S$ and we have introduced the one dimensional projections
$\Pi_{\varphi_i} = | \varphi_i \rangle \langle \varphi_i |$. The statistical
operator $W$, which is not a pure state, admits many different decompositions.
In particular one can consider an orthogonal decomposition
\begin{eqnarray}
  W & = & \sum_{i = d}^n w_i | \varphi_i \rangle \langle \varphi_i |,
  \label{eq:ort} 
\end{eqnarray}
given by its spectral resolution, where the $\{ \varphi_i \}_{i = d, \ldots
n}$ are orthonormal vectors, further orthogonal to the span of $\{ \varphi_i
\}_{i = 1, \ldots, d - 1}$, so that altogether they provide a basis in
$\mathcal{H}_S$, and the positive weights $w_i$ sum up to one. At the same
time one can consider many others non orthogonal convex decompositions of the
form
\begin{eqnarray}
  W & = & \sum_{k = 1}^r \mu_k | \psi_k \rangle \langle \psi_k |
  \label{eq:nort}, 
\end{eqnarray}
with $r > n - (d - 1)$ and where the $\{ \psi_k \}_{k = 1, \ldots, r}$ are
normalized but generally non orthogonal states while $\{ \mu_k \}_{k = 1,
\ldots, r}$ is a probability distribution. For a choice of system states of
the form Eq.~(\ref{eq:ssi}) the compatibility domain Eq.~(\ref{eq:cs}) can
therefore be seen to arise from the two following distinct sets of correlated
system environment states, namely
\begin{eqnarray}
  \mathcal{C}_{SE}^I & = & \left\{ \rho_{SE} = \sum_{i = 1}^{d - 1} p_i |
  \varphi_i \rangle \langle \varphi_i | \otimes \rho^i_E + p_d \sum^n_{i = d}
  w_i | \varphi_i \rangle \langle \varphi_i | \otimes \varrho^i_E \right\}
  \label{eq:c0} 
\end{eqnarray}
and
\begin{eqnarray}
  \mathcal{C}^{II}_{SE} & = & \left\{ \rho_{SE} = \sum_{i = 1}^{d - 1} p_i |
  \varphi_i \rangle \langle \varphi_i | \otimes \rho^i_E + p_d \sum_{k = 1}^r
  \mu_k | \psi_k \rangle \langle \psi_k | \otimes \varrho^k_E \right\}
  \label{eq:c1}, 
\end{eqnarray}
with $\sum_{i = 1}^d p_i = 1$, while $\{ \rho_E^i \}_i$ and $\{ \varrho_E^k
\}_k$ are collections of distinct environmental states. While these composite
states have the same compatibility domain according to Eq.~(\ref{eq:ssi}), we
have the important difference that while $\mathcal{C}_{SE}^I$ only contains
zero quantum discord states, this is no more true for $\mathcal{C}^{II}_{SE}$,
which thus also includes quantum correlations. Given an arbitrary system
environment interaction $U_{SE} (t)$ we can now consider the transformation
$\Phi^{II} (t)$ which associates to the marginal of a state $\rho_{SE} \in
\mathcal{C}_{SE}^{II}$, that is
\begin{eqnarray}
  \rho_S (0) & = & \tmop{Tr}_E \rho_{SE} \nonumber\\
  & = & \sum_{i = 1}^{d - 1} p_i | \varphi_i \rangle \langle \varphi_i | +
  p_d W, \label{eq:margi} 
\end{eqnarray}
the marginal associated to the time evolved state according to
\begin{eqnarray}
  \rho_S (0) & \rightarrow & \rho_S (t) = \tmop{Tr}_E (U_{SE} (t) \rho_{SE}
  U^{\dag}_{SE} (t)), \label{eq:rst} 
\end{eqnarray}
so that we set
\begin{eqnarray}
  \rho_S (t) & = & \Phi^{II} (t) [\rho_S (0)], \label{eq:map2} 
\end{eqnarray}
and an analogue construction can be done starting from states in
$\mathcal{C}_{SE}^I$, thus obtaining a collection of maps $\Phi^I (t)$. We
have now the important fact that such assignments actually define positive
affine maps on the convex set $\mathcal{C}_S$, which can be uniquely extended
to linear maps on the linear hull of $\mathcal{C}_S$. Since the elements of
the set $\mathcal{C}_S$ commute, according to
{\cite{Stinespring1955a,Takesaki2002}} we therefore have that such maps are
actually completely positive. We can now build on another fundamental result
about linear maps which are completely positive, namely the fact that they can
be expressed in the so-called Kraus form {\cite{Kraus1983a}}.

To explicitly exhibit a Kraus representation for the considered maps we
proceed as follows. Let us first evaluate the trace in Eq.~(\ref{eq:rst}) by
considering a complete orthonormal system $\{ | \gamma \rangle \nobracket \}$
in $\mathcal{H}_E$, thus obtaining
\begin{eqnarray}
  \rho_S (t) & = & \sum_{i = 1}^{d - 1} \sum_{\alpha_i, \gamma}
  \lambda_{\alpha_i} \langle \gamma | U_{SE} (t) | \alpha_i \rangle p_i |
  \varphi_i \rangle \langle \varphi_i | \langle \alpha_i | U_{SE}^{\dag} (t) |
  \gamma \rangle \nonumber\\
  &  & + \sum_{k = 1}^r \sum_{\beta_k, \gamma} \eta_{\beta_k} \langle \gamma
  | U_{SE} (t) | \beta_k \rangle p_d \mu_k | \psi_k \rangle \langle \psi_k |
  \langle \beta_k | U^{\dag}_{SE} (t) | \gamma \rangle \label{eq:vk} 
\end{eqnarray}
where we have also introduced orthogonal decompositions for the environmental
operators appearing in Eq.~(\ref{eq:c0}) and Eq.~(\ref{eq:c1}) according to
$\rho^i_E = \sum_{\alpha_i} \lambda_{\alpha_i} | \alpha_i \rangle \langle
\alpha_i |$ and \ $\varrho^k_E = \sum_{\beta k} \eta_{\beta_k} | \beta_k
\rangle \langle \beta_k |$. We now want to recast Eq.~(\ref{eq:vk}) as a
linear action on $\rho_S (0)$ as given by Eq.~(\ref{eq:margi}). To this aim we
first observe that we have
\begin{eqnarray}
  \Pi_{\varphi_i} \rho_S (0) \Pi_{\varphi_i} & = & p_i | \varphi_i \rangle
  \langle \varphi_i | \label{eq:easy}, 
\end{eqnarray}
which allows us to express in the desired fashion the first line of
Eq.~(\ref{eq:vk}). To proceed further we exploit a general theorem which
connects different possible orthogonal and non orthogonal decompositions of a
given quantum state. The theorem was first formulated by Schr{\"o}dinger
{\cite{Schrodinger1935a}} and later rediscovered by Gisin {\cite{Gisin1989a}}
as well as Hughston, Josza and Wootters {\cite{Hughston1993a}}, so that it is
often known as GHJW theorem. According to this theorem given the two
decompositions Eq.~(\ref{eq:ort}) and Eq.~(\ref{eq:nort}) of the statistical
operator $W$ there exists a unitary matrix $U$, whose columns are given by
$U_{kj} = \sqrt{\mu_k / w_j} \langle \varphi_j | \psi_k \rangle \nobracket$
for $j = d, \ldots, n$, so that in particular setting
\begin{eqnarray}
  \lambda_{kj} & = & | U_{kj} |^2 \nonumber\\
  & = & \frac{\mu_k}{w_j} | \langle \varphi_j | \psi_k \rangle \nobracket |^2
  \label{eq:u} 
\end{eqnarray}
we have for all $k$
\begin{eqnarray}
  \mu_k & = & \sum_{j = d}^n \lambda_{kj} w_j . \label{eq:link} 
\end{eqnarray}
We can thus introduce the operators
\begin{eqnarray}
  K_{jk} & = & \sqrt{\lambda_{kj}} | \psi_k \rangle \langle \varphi_j |,
  \label{eq:K} 
\end{eqnarray}
satisfying the relation
\begin{eqnarray}
  \sum^n_{j = d} K_{jk} \rho_S (0) K_{jk}^{\dag} & = & p_d \mu_k | \psi_k
  \rangle \langle \psi_k |, \label{eq:inv} 
\end{eqnarray}
which allows to express the second line of Eq.~(\ref{eq:vk}) as a linear
trasformation acting on $\rho_S (0)$. Thanks to Eq.~(\ref{eq:easy}) and
Eq.~(\ref{eq:inv}) we can finally introduce the system operators
\begin{eqnarray}
  M_{\gamma \alpha_i} (t) & = & \sqrt{\lambda_{\alpha_i}} \langle \gamma |
  U_{SE} (t) | \alpha_i \rangle \Pi_{\varphi_i} \label{eq:m1} 
\end{eqnarray}
together with
\begin{eqnarray}
  M^j_{\gamma \beta_k} (t) & = & \sqrt{\eta_{\beta_k}} \langle \gamma | U_{SE}
  (t) | \beta_k \rangle K_{jk}, \label{eq:m2} 
\end{eqnarray}
which provide an explicit Kraus representation of the map $\Phi^{II} (t)$
defined through Eq.~(\ref{eq:vk})
\begin{eqnarray}
  \rho_S (t) & = & \sum_{i = 1}^{d - 1} \sum_{\gamma \nocomma, \alpha_i}
  M_{\gamma \alpha_i} (t) \rho_S (0) M_{\gamma \alpha_i} (t)^{\dag} + \sum_{k
  = 1}^r \sum^n_{j = d} \sum_{\gamma, \beta_k} M^j_{\gamma \beta k} (t) \rho_S
  (0) M^j_{\gamma \beta_k} (t)^{\dag} . \label{eq:Kraus} 
\end{eqnarray}
The obtained expression for the map $\Phi^{II} (t)$ allows to extend it by
linearity to the whole set of system states $\mathcal{S} (\mathcal{H}_S)$ in
Kraus form, thus remaining completely positive. This construction contains as
special case the examples considered in {\cite{Brodutch2013a,Liu2014a}}.

Note that through this construction besides the collection of time dependent
positive operator-valued measures naturally associated to the family of
channels \ $\Phi^{II} (t)$ thanks to trace preservation {\cite{Nielsen2000}},
one can also put into evidence a positive operator-valued measure determined
by the class of correlated system-environment states. The latter is given by
the set $\{ \Pi_{\varphi_i}, K_{jk}^{\dag} K_{jk} \}_{i, j, k}$, where the
indexes take on the values $i = 1, \ldots, d - 1$, $j = d, \ldots, n$ and $k =
1, \ldots, r$. It is actually fixed by the following transformation which
leaves invariant the compatibility domain associated to
$\mathcal{C}^{II}_{SE}$
\begin{eqnarray}
  \sum_{i = 1}^{d - 1} \Pi_{\varphi_i} \rho_S (0) \Pi_{\varphi_i} + \sum_{k =
  1}^r \sum_{j = d}^n K_{jk} \rho_S (0) K^{\dag}_{jk} & = & \rho_S (0),
  \label{eq:inva} 
\end{eqnarray}
with $\rho_S (0)$ as in Eq.~(\ref{eq:margi}). The analogous construction for
the family of channels $\Phi^I (t)$, leads in particular to the
projection-valued measure $\{ \Pi_{\varphi_i} \}_i$, with $i = 1, \ldots, n$.

We have thus provided a general construction to define a completely positive
map starting from a correlated system-environment state, which generally has
non zero quantum discord, belonging to a convex subset of the whole set of
states whose compatibility domain is actually a commutative set. In such a way
for arbitrary system-environment interaction one obtains a positive map
providing the dynamical evolution of the reduced system, which due to
commutativity of the domain on which it is defined, namely the linear hull of
the convex compatibility domain, has to be completely positive and therefore
admits a Kraus representation. Given its expression in terms of Kraus
operators the map can be extended as completely positive map to the whole set
of reduced system states. As we shall stress below, this extension is in
general highly non unique. Furthermore the very construction of the maps do
depend on both the reduced system state, the environmental states and the
particular correlations. In the case of a reduced system state with a
degenerate spectrum in particular the same state can belong to compatibility
domains arising from different correlated states, and system-environment
states with the very same marginals lead to utterly different maps.

\

\subsection{Zero quantum discord states and non uniqueness of the
construction}

\begin{figure}[h]
\label{fig:2maps}
 \centering
\includegraphics[width=\linewidth]{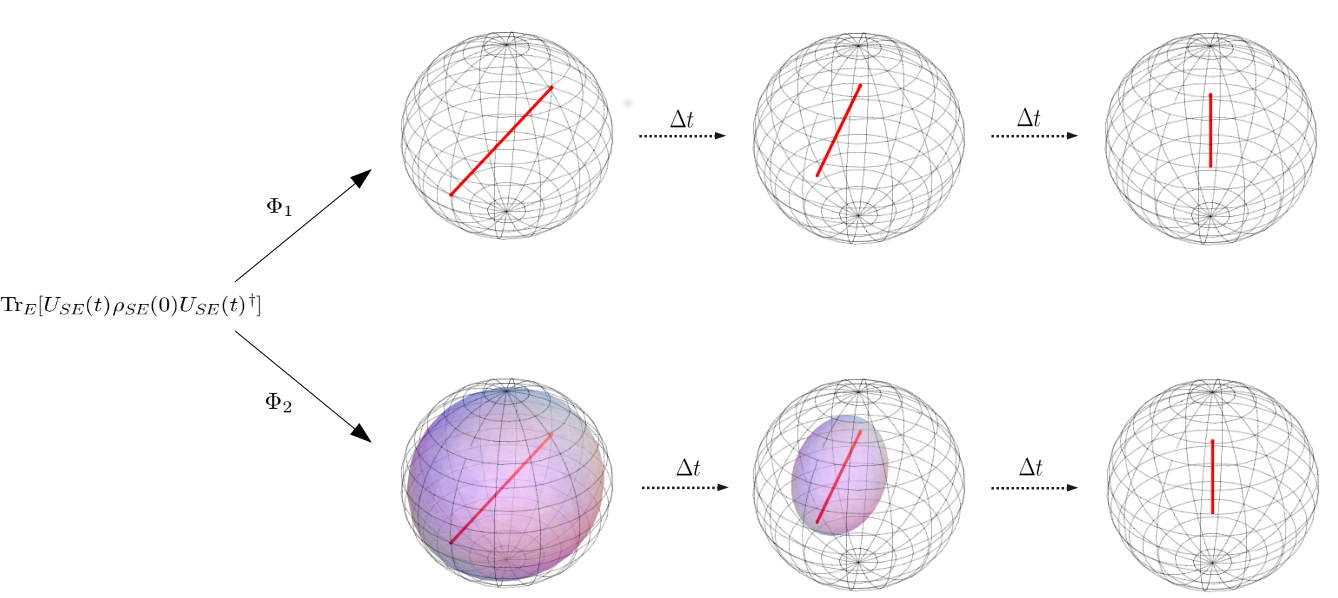}
  \caption{Schematic illustration of the different action of the maps $\Phi_1
  (t)$ and $\Phi_2 (t)$ defined in Eq.~(\ref{eq:K1}) and Eq.~(\ref{eq:K2})
  respectively. The different plots show the transformed Bloch sphere at
  subsequent times for the choice $y = \frac{1}{5}$ and $\chi = - \frac{2}{5}$
  in Eq.~(\ref{eq:stato0qd}). The commutative compatibility domain is here
  given by a diameter of the sphere, which transforms in the same way under
  the action of the maps, while the transformation of the rest of the sphere
  does depend on the choice of extension.}
\end{figure}

A similar but simpler construction with respect to the one considered above
can be obtained for zero quantum discord states, starting from the set
$\mathcal{C}^I_{SE}$, in analogy with the result obtained in
{\cite{Rodriguez2008a}}. We will consider this situation to put into evidence
the non uniqueness of the completely positive extension of the maps initially
defined only on the compatibility domain made up of commuting statistical
operators for the system, a point which has actually not yet been put into
evidence. Indeed considering a system statistical operator in the
compatibility domain associated to $\mathcal{C}^I_{SE}$, namely of the form
\begin{eqnarray}
  \rho_S (0) & = & \tmop{Tr}_E \rho_{SE} \nonumber\\
  & = & \sum_{i = 1}^{d - 1} p_i | \varphi_i \rangle \langle \varphi_i | +
  p_d \sum^n_{i = d} w_i | \varphi_i \rangle \langle \varphi_i | \nonumber\\
  & = & \sum_{i = 1}^n \tilde{p}_i | \varphi_i \rangle \langle \varphi_i |
  \label{eq:margi2}, 
\end{eqnarray}
we can introduce two channels sending states in $\mathcal{S} (\mathcal{H}_S)$
to states in $\mathcal{S} (\mathcal{H}_S \otimes \mathcal{H}_E)$, namely
\begin{eqnarray}
  \mathcal{A}_1 [\sigma] & = & \sum_{j = 1}^n \Pi_{\varphi_j} \sigma
  \Pi_{\varphi_j} \otimes \rho_E^j \label{eq:a1} 
\end{eqnarray}
and
\begin{eqnarray}
  \mathcal{A}_2 [\sigma] & = & \sum_{j, k = 1}^n \sum_{\alpha_j, \alpha_k}
  \sqrt{\lambda_{\alpha_j} \lambda_{\alpha_k}} \Pi_{\varphi_j} \sigma
  \Pi_{\varphi_k} \otimes | \alpha_j \rangle \langle \alpha_k |,^j
  \label{eq:a2} 
\end{eqnarray}
which coincide on states belonging to the compatibility domain. The two
channels are written in Kraus form, so that they can be extended from the
linear hull of the compatibility domain to the whole linear space of trace
class operators on $\mathcal{H}_S$, thus allowing to define the completely
positive maps
\begin{eqnarray}
  \Phi_{1, 2} (t) & = & \tmop{Tr}_E \circ \mathcal{U}_{SE} (t) \circ
  \mathcal{A}_{1, 2}, \label{eq:f1} 
\end{eqnarray}
where $\mathcal{U}_{SE} (t) [\sigma] = U_{SE} (t) \sigma U^{\dag}_{SE} (t)$.
We note in particular that upon introducing the diagonalizing projection
\begin{eqnarray}
  \Phi_d [w] & = & \sum_{j = 1}^n \Pi_{\varphi_j} w \Pi_{\varphi_j}
  \label{eq:fd} 
\end{eqnarray}
we have
\begin{eqnarray}
  \Phi_1 (t) & = & \Phi_2 (t) \circ \Phi_d . \label{eq:f2} 
\end{eqnarray}
Considering the action of $\mathcal{A}_1$ on states of the form
Eq.~(\ref{eq:margi2}) we obtain a representation of $\Phi_1 (t)$ in Kraus form
as
\begin{eqnarray}
  \Phi_1 (t) [\rho_S (0)] & = & \sum_{i = 1}^n \sum_{\gamma \nocomma,
  \alpha_i} M_{\gamma \alpha_i} (t) \rho_S (0) M_{\gamma \alpha_i} (t)^{\dag}
  \label{eq:K1}, 
\end{eqnarray}
which is the analogue of Eq.~(\ref{eq:Kraus}) for states coming from
$\mathcal{C}^I_{SE}$. We can however also consider the channel map
$\mathcal{A}_2$ and thus come to the completely positive map
\begin{eqnarray}
  \Phi_2 (t) [\rho_S (0)] & = & \sum_{\gamma \nocomma, \tmmathbf{\alpha}}
  M_{\gamma \tmmathbf{\alpha}} (t) \rho_S (0) M_{\gamma \tmmathbf{\alpha}}
  (t)^{\dag} \label{eq:K2}, 
\end{eqnarray}
where we have defined another set of Kraus operators according to
\begin{eqnarray}
  M_{\gamma \tmmathbf{\alpha}} (t) & = & \sum_{i = 1}^n M_{\gamma \alpha_i}
  (t) . \label{eq:K3} 
\end{eqnarray}
The particular representation Eq.~(\ref{eq:K2}) corresponds to the theorem
considered in {\cite{Rodriguez2008a}}, even though, as shown in the example
below, in the paper actually a different expression was in fact used
to obtain a completely positive map starting from a zero discord correlated
state and a given dynamics. Note that both $\Phi_1 (t)$ and $\Phi_2 (t)$
coincide when applied to states belonging to the compatibility domain, that is
in this case diagonal in the basis $\{ | \varphi_i \rangle \nobracket \}_i$,
while differ in their action on the rest of the states. This situation is
schematically depicted in Fig.~(\ref{fig:2maps}) with reference to the example
treated below. This difference is actually amenable to experimental
observation, and indeed the different evolution in time through reduced maps
connected via the relation Eq.~(\ref{eq:f2}) has been exploited to
experimentally detect initial correlations between system and environment
{\cite{Gessner2011a,Gessner2013a,Gessner2014a,Cialdi2014a}}. The key point
lies in the extension of the map from the initial domain to the whole set of
states, which as we have shown can be performed in different ways. It is to be
stressed that, besides acting differently on states outside the compatibility
domain, the maps do generally not reduce to the identity for $t = 0$. This
corresponds to the fact that the channels defined in Eq.~(\ref{eq:a1}) and
Eq.~(\ref{eq:a2}), when composed with the partial trace with respect to the
environmental degrees of freedom, act as the identity only on the
compatibility domain, so that they do not define proper assignment maps
{\cite{Pechukas1994a,Alicki1995a}}.

\subsection{Correlated qubit states}

The non uniqueness of the proposed constructions can be better seen
considering the following example. Let both $\mathcal{H}_S$ and
$\mathcal{H}_E$ be isomorphic to $\mathbbm{C}^2$, and consider the correlated
zero quantum discord state
\begin{eqnarray}
  \rho_{SE} & = & \sum_{i = 1, 2} p_i \Pi_S^i \otimes \rho_E^i .
  \label{eq:cesar} 
\end{eqnarray}
Denoting with $\{ \sigma^0, \sigma^1, \sigma^2, \sigma^3 \} \equiv \{
\mathbbm{1}, \sigma_x, \sigma_y, \sigma_z \}$ the basis of linear operators on
$\mathbbm{C}^2$ made up of the identity and the Pauli matrices, we take $\{
\Pi_S^i \}_{i = 1, 2}$ to be the projections on the eigenvectors of $\sigma_y$
and assume $\{ \rho^i_E \}_{i = 1, 2}$ to be diagonal in the computational
basis determined by the eigenvectors of $\sigma_z$ corresponding to the
eigenvalues $+ 1$ $(| 0 \rangle \nobracket)$ and $- 1$ $(| 1 \rangle
\nobracket)$, according to
\begin{eqnarray}
  \rho^i_E & = & \lambda_{\noplus \noplus + i} | 0 \rangle \langle 0 | +
  \lambda_{\noplus \noplus - i} | 1 \rangle \langle 1 |, \nonumber
\end{eqnarray}
with $\lambda_{\noplus \noplus + i} + \lambda_{\noplus \noplus - i} = 1$. Let
us further consider a unitary system environment interaction of the form
\begin{equation}
  U_{SE} (t) = \prod_{j = 1}^3 [\cos (\omega t) \mathbbm{1}_S \otimes
  \mathbbm{1}_E - i \sin (\omega t) \sigma_S^j \otimes \sigma_E^j],
  \label{eq:unitiniz}
\end{equation}
which allows for an analytic evaluation of the time evolution maps. The Choi
matrices associated to these mappings can be easily obtained exploiting the
relation Eq.~(\ref{eq:mchoi}). Neglecting time arguments for simplicity they
can be compactly written
\begin{equation}
  \Phi^1_{\tmop{Choi}} = \frac{1}{2} \left(\begin{array}{cccc}
    C^2 + 2 S^2 \mu_+ & 2 SC (\varkappa_- - \varkappa_+) & 4 iS^2 \varkappa_+
    & \begin{array}{c}
      C^2 - iCS (\mu_+ - \mu_-)
    \end{array}\\
    2 SC (\varkappa_- - \varkappa_+) & C^2 + 2 S^2 \mu_- & - C^2 - iCS (\mu_+
    - \mu_-) & 4 iS^2 \varkappa_-\\
    - 4 iS^2 \varkappa_+ & - C^2 + iCS (\mu_+ - \mu_-) & C^2 + 2 S^2 \mu_+ & 2
    SC (\varkappa_- - \varkappa_+)\\
    C^2 + iCS (\mu_+ - \mu_-) & - 4 iS^2 \varkappa_- & 2 SC (\varkappa_- -
    \varkappa_+) & C^2 + 2 S^2 \mu_-
  \end{array}\right) \label{eq:Choiphi1}
\end{equation}
together with
\begin{equation}
  \Phi^2_{\tmop{Choi}} = \left(\begin{array}{cccc}
    C^2 \chi_+ + S^2 \mu_+ & SC (\varkappa_- - \varkappa_+) & 2 iS^2
    \varkappa_+ & \begin{array}{c}
      C^2 \chi_+ - iCS \chi_-
    \end{array}\\
    SC (\varkappa_- - \varkappa_+) & C^2 \kappa_+ + S^2 \mu_- & - C^2 \kappa_+
    - iCS \kappa_- & 2 iS^2 \varkappa_-\\
    - 2 iS^2 \varkappa_+ & - C^2 \kappa_+ + iCS \kappa_- & C^2 \kappa_+ + S^2
    \mu_+ & SC (\varkappa_- - \varkappa_+)\\
    \begin{array}{c}
      C^2 \chi_+ + iCS \chi_-
    \end{array} & - 2 iS^2 \varkappa_- & SC (\varkappa_- - \varkappa_+) & C^2
    \chi_+ + S^2 \mu_-
  \end{array}\right) \label{eq:Choiphi2},
\end{equation}
where we have denoted $C (t) = \cos (2 \omega t)$, $S (t) = \sin (2 \omega t)$
and the different constants are functions of the eigenvalues of the
environmental statistical operators as detailed in the Appendix, see
Eq.~(\ref{eq:label}).

It is interesting to compare these results with the analysis performed in
{\cite{Rodriguez2008a}}, where the problem of obtaining a completely positive
mapping starting from a correlated state was addressed for zero discord
states. In particular the authors considered the state
\begin{equation}
  \rho_{SE} = \frac{1}{4}  [\mathbbm{1}_S \otimes \mathbbm{1}_E + y \sigma_S^2
  \otimes \mathbbm{1}_E - \chi \sigma_S^2 \otimes \sigma_E^3],
  \label{eq:stato0qd}
\end{equation}
which we recover upon setting
\begin{eqnarray}
  \{ p_1, \lambda_{+ 1}, \lambda_{- 1} \} & \rightarrow & \left\{ \frac{1}{2},
  \frac{1}{2} \left( 1 - \frac{\chi}{1 + y} \right), \frac{1}{2} \left( 1 +
  \frac{\chi}{1 + y} \right) \right\} \nonumber\\
  \{ p_2, \lambda_{+ 2}, \lambda_{- 2} \} & \rightarrow & \left\{ \frac{1}{2},
  \frac{1}{2} \left( 1 + \frac{\chi}{1 - y} \right), \frac{1}{2} \left( 1 -
  \frac{\chi}{1 - y} \right) \right\}, \label{eq:valori} 
\end{eqnarray}
where the constraints $| y | \leqslant 1$ and $\chi^2 \leqslant 1 - y^2$ hold,
and let it evolve according to Eq.~(\ref{eq:unitiniz}). In particular by
substituting the values Eq.~(\ref{eq:valori}) in the expression
Eq.~(\ref{eq:label}) one can obtain from Eq.~(\ref{eq:Choiphi2}) the
corresponding expression for the Choi matrix, given in Eq.~(\ref{eq:mjpa}) of
the Appendix. It can be noted that this matrix does not coincide with
the result presented in Sec. 5 of {\cite{Rodriguez2008a}}, despite the fact
that the authors there advocate just one of the constructions that we used to
come from a zero quantum discord state to the completely positive map, namely
Eq.~(\ref{eq:K2}). The result can be understood as follows. The composition of
any positive map, coming e.g. from a reduced dynamics as in
Eq.~(\ref{eq:map}), with a diagonalizing map as in Eq.~(\ref{eq:fd}) leads to
a completely positive map. Indeed by composing a map with a diagonalizing map
we obtain for the map a commutative domain, on which positivity is equivalent
to complete positivity, independently of the choice of orthogonal projections
$\{ \Pi_{\varphi_i} \}_i$.

\subsection{Reduced dynamics from discordant states}

As simplest example of a completely positive map obtained starting from
discordant states let us \ consider $\mathcal{H}_S$ of dimension $n$ and take
the state
\begin{eqnarray}
  \rho_{SE} & = & \sum_{i = 1}^{n - 2} p_i | \varphi_i \rangle \langle
  \varphi_i | \otimes \rho^i_E + p_d \sum_{k = 1}^3 \mu_k | \psi_k \rangle
  \langle \psi_k | \otimes \varrho_E^k \label{eq:again} 
\end{eqnarray}
where introducing the two orthonormal states $\{ | 0 \rangle \nobracket, | 1
\rangle \nobracket \}$, further orthogonal to $\{ | \varphi_i \rangle
\nobracket \}_{i = 1, \ldots, d}$, we have $\left\{ | \psi_1 \rangle
\nobracket = | 0 \rangle \nobracket, | \psi_2 \rangle \nobracket = | 1 \rangle
\nobracket, | \psi_3 \rangle \nobracket = | + \rangle \right\}$, with $| +
\rangle \nobracket = (| 0 \rangle \nobracket + | 1 \rangle \nobracket) /
\sqrt{2}$ and $| - \rangle \nobracket = (| 0 \rangle \nobracket - | 1 \rangle
\nobracket) / \sqrt{2}$. The state $W$ can therefore be written
\begin{eqnarray}
  W & = & \left(\begin{array}{cc}
    \mu_1 + \frac{\mu_3}{2} & \frac{\mu_3}{2}\\
    \frac{\mu_3}{2} & \mu_2 + \frac{\mu_3}{2}
  \end{array}\right), \label{eq:W} 
\end{eqnarray}
and one can consider its spectral decomposition
\begin{eqnarray}
  W & = & \sum_{i = 1, 2} w_i | \phi_i \rangle \langle \phi_i |,
  \label{eq:Wdiag} 
\end{eqnarray}
leading according to Eq.~(\ref{eq:K}) to a set of six Kraus operators. For the
simplest case of a uniform probability distribution, so that $\mu_k =
\frac{1}{3}$ for all $k$, we have $\{ | \phi_1 \rangle \nobracket = | +
\rangle \nobracket, | \phi_2 \rangle \nobracket = | - \rangle \nobracket \}$
together with $\{ w_1 = 2 / 3, w_2 = 1 / 3 \}$. This choice of parameters
allows us to make direct contact with the example considered in
{\cite{Brodutch2013a}}. We obtain in particular the set of Kraus operators
$K_{jk} = \sqrt{\lambda_{kj}} | \psi_k \rangle \langle \phi_j |$
\begin{eqnarray}
  K_{+ 0} = \sqrt{\frac{1}{4}} |0 \rangle \langle + | \qquad K_{+ 1} =
  \sqrt{\frac{1}{4}} |1 \rangle \langle + | \qquad K_{+ +} =
  \sqrt{\frac{1}{2}} | + \rangle \langle + | &  & \label{eq:POVMBr} \\
  K_{- 0} = \sqrt{\frac{1}{2}} |0 \rangle \langle - | \qquad K_{- 1} =
  \sqrt{\frac{1}{2}} |1 \rangle \langle - | \qquad K_{- +} = 0 &  & 
\end{eqnarray}
which leave $W$ defined in Eq.~(\ref{eq:Wdiag}) invariant according to
\[ \sum_{k = 1, 2, 3} \sum_{j = \pm} K_{jk} WK_{jk}^{\dagger} = W, \]
and leading according to the general theory to the positive operator-valued
measure $\{\Pi_{\varphi_i}, K_{jk}^{\dagger} K_{jk} \}_{i = 1, \cdots, n - 2 ;
j = \pm ; k = 1, 2, 3}$. Note that this set of Kraus operators does not
coincide with those exhibited in {\cite{Brodutch2013a}}. This fact can again
be traced back to the non uniqueness in the construction of the completely
positive map. Indeed while the action of the map on the set of operators
commuting with the marginal of Eq.~(\ref{eq:again}) is uniquely defined, the
extension to the whole space of statistical operators can be obtained in many
ways, still preserving complete positivity. In particular it can be seen that
the set $\{ K_{jk} \}$ with $K_{jk} = \sqrt{\mu_k} | \chi_j \rangle \langle
\psi_k |$ still obeys Eq.~(\ref{eq:inv}) for any collection $\{ \chi_j \}_j$
of normalized but not necessarily orthogonal states such that $\sum_j | \chi_j
\rangle \langle \chi_j |$ acts as the identity on the space spanned by $\{
\phi_i \}_{i = 1, 2}$.

\section{Discussion}

The possibility to consider a reduced dynamical description for a given set of
quantum degrees of freedom interacting with some external environment provides
a very convenient way to account for the observed dynamics of such degrees of
freedom. In this respect open quantum system theory has led to a satisfactory
explanation of various physical phenomena and its general framework provides
viable schemes to cope with the description of dissipation and decoherence
effects in many different fields. However according to the general theory it
is clear how to obtain a reduced dynamics only in the presence of an initially
factorized system-environment state, a condition which cannot always be
considered as realistic in the presence of strong coupling. The extension of
the formalism to include correlated initial states however appears to be non
trivial and not always bears with itself the desired properties. In this
article we have provided a general construction of dynamical map, for an
arbitrary unitary interaction between system and environment, for a class of
correlated states possibly including states with non zero discord. This result
encompasses previous work and put it within a unified viewpoint. It also shows
that the definition of a reduced dynamical map in the case of correlated
states is linked to the introduction of a set of Kraus operators building up a
positive operator-valued measure which leaves the reduced system state
invariant. The key observation lies in the characterization of the set of
reduced states compatible with given correlations. If this compatibility
domain is made up of commuting states, exploiting the identification between
positivity and complete positivity on such sets one can actually introduce
well defined evolution maps. The latter can also be extended to the whole set
of statistical operators, still retaining the property of complete positivity.
However such extensions are generally highly non unique, as we have explicitly
pointed out by means of example. This point, to the bet of our knowledge, has
yet not been put into the due evidence in the literature, and has allowed us
to better clarify previous special results
{\cite{Rodriguez2008a,Brodutch2013a}}. Indeed while coinciding in their action
on the compatibility domain, they generally transform in a different way
states outside this domain. Moreover outside the compatibility domain they do
not necessarily act as the identity at the initial time, thus describing a
kind of initial slippage. The obtained picture, while elucidating a few basic
points, and providing a constructive approach, further shows that extension of
such maps beyond their natural domain, while preserving complete positivity is
not necessarily of direct physical relevance.

It remains an open and relevant question whether the formalism can also be
extended to states containing quantum correlations in the form of
entanglement.

\appendix

\section{Construction of the mappings $\Phi_1$ and $\Phi_2$}

We now consider how to explicitly obtain the completely positive maps $\Phi_1
(t)$ and $\Phi_2 (t)$ starting from the correlated state considered in
Eq.~(\ref{eq:cesar}), according to the dynamics described by
Eq.~(\ref{eq:unitiniz}). In order to identify the completely positive maps
Eq.~(\ref{eq:K1}) and Eq.~(\ref{eq:K2}) we exploit a matrix representation of
these maps {\cite{Heinosaari2011}}, given by
\begin{eqnarray}
  \Lambda^{1, 2}_{ij, kl} (t) & = & \langle i | \Phi^{}_{1, 2} (t) [| k
  \rangle \langle l |] | j \rangle, \label{eq:mmap} 
\end{eqnarray}
where the indexes take on the values 0 and 1. To actually evaluate the matrix
elements we observe that the unitary evolution given by
Eq.~(\ref{eq:unitiniz}) can be written, up to an irrelevant phase factor, in
the form
\begin{equation}
  U_{SE} (t) = \cos (2 \omega t)  \mathbbm{1}_S \otimes \mathbbm{1}_E -
  \frac{i}{2} \sin (2 \omega t) \sum_{j = 0}^3 \sigma_S^j \otimes \sigma_E^j .
  \label{eq:unit}
\end{equation}
A straightforward but lengthy calculation then leads to the explicit
expression for the matrices $\Lambda^{1, 2}_{ij, kl} (t)$, which act
identically on states diagonal in the eigenbasis of $\sigma_y$, while
transforming in a different way system states diagonal in different bases. In
particular they generally do not act as the identity map for $t = 0$.

We start considering the matrix associated to $\Phi_2 (t)$. We need to
evaluate the operators
\begin{eqnarray}
  M_{\gamma \tmmathbf{\alpha}} (t) & = & \sum_{i = 1, 2} M_{\gamma \alpha_i}
  (t) \label{eq:m} \\
  & = & \sum_{i = 1, 2} \sqrt{\lambda_{\alpha_i}}  \langle \gamma |U_{SE} (t)
  | \alpha_i \rangle \Pi_{\varphi_i}, \nonumber
\end{eqnarray}
where $\{ \Pi_{\varphi_i} \}_i$ now denote the projections $\{ \Pi_S^i \}_{i =
1, 2}$ on the eigenvectors of $\sigma_y$ and $\{ | \alpha_i \rangle \nobracket
\}_i$ are actually independent from $i$, since both environmental statistical
operators are diagonal in the computational basis. We obtain
\begin{eqnarray}
  \langle 0| U_{SE} (t) |0 \rangle & = & \cos (2 \omega t)  \mathbbm{1}_S - i
  \sin (2 \omega t) | 0 \rangle \langle 0 | \nonumber\\
  \langle 0| U_{SE} (t) |1 \rangle & = & - i \sin (2 \omega t) \sigma_S^-
  \nonumber\\
  \langle 1| U_{SE} (t) |0 \rangle & = & - i \sin (2 \omega t) \sigma_S^+
  \nonumber\\
  \langle 1| U_{SE} (t) |1 \rangle & = & \cos (2 \omega t)  \mathbbm{1}_S - i
  \sin (2 \omega t) | 1 \rangle \langle 1 |, \label{eq:uelem} 
\end{eqnarray}
where we have denoted $C (t) = \cos (2 \omega t)$, $S (t) = \sin (2 \omega t)$
and defined the raising and lowering operators according to $\sigma_S^+ = |0
\rangle \langle 1|$ and $\sigma_S^- = |1 \rangle \langle 0|$, leading
according to Eq.~(\ref{eq:K3}) to
\begin{eqnarray}
  M_{000} & = & C (\sqrt{\lambda_{+ 1}} \Pi_S^1 + \sqrt{\lambda_{+ 2}}
  \Pi_S^2) - iS \Sigma_+  |0 \rangle \langle 0| - S \Delta_+ \sigma_S^+
  \nonumber\\
  M_{011} & = & - iS \Sigma_- \sigma_S^- - S \Delta_-  |1 \rangle \langle 1|
  \nonumber\\
  M_{100} & = & - iS \Sigma_+ \sigma_S^+ + S \Delta_+  |0 \rangle \langle 0|
  \nonumber\\
  M_{111} & = & C (\sqrt{\lambda_{- 1}} \Pi_S^1 + \sqrt{\lambda_{- 2}}
  \Pi_S^2) - iS \Sigma_-  |1 \rangle \langle 1| + S \Delta_- \sigma_S^-
  \label{eq:kraus2}, 
\end{eqnarray}
and used the notation $M_{\gamma \tmmathbf{\alpha}} (t) \rightarrow M_{\gamma
\alpha_1 \alpha_2}$. One can directly check the identity $\sum_{\gamma
\nocomma, \tmmathbf{\alpha}} M_{\gamma \tmmathbf{\alpha}} (t)^{\dag} M_{\gamma
\tmmathbf{\alpha}} (t) = \mathbbm{1}_S$, granting trace preservation of the
map. Computing the action of the map on the computational basis by evaluating
$\sum_{\gamma, \alpha} M_{\gamma \tmmathbf{\alpha}} (t) |k \rangle \langle
l|M_{\gamma \tmmathbf{\alpha}} (t)^{\dagger}$ and taking the matrix elements
one can find the matrix $\Lambda^2_{ij, kl} (t)$
\begin{equation}
  \Lambda^2 = \left(\begin{array}{cccc}
    C^2 \chi_+ + S^2 \mu_+ & SC (\varkappa_- - \varkappa_+) & SC (\varkappa_-
    - \varkappa_+) & C^2 \kappa_+ + S^2 \mu_-\\
    2 iS^2 \varkappa_+ & \begin{array}{c}
      C^2 \chi_+ - iCS \chi_-
    \end{array} & \begin{array}{c}
      - C^2 \kappa_+ - iCS \kappa_-
    \end{array} & 2 iS^2 \varkappa_-\\
    - 2 iS^2 \varkappa_+ & - C^2 \kappa_+ + iCS \kappa_- & \begin{array}{c}
      C^2 \chi_+ + iCS \chi_-
    \end{array} & - 2 iS^2 \varkappa_-\\
    \begin{array}{c}
      C^2 \kappa_+ + S^2 \mu_+
    \end{array} & SC (\varkappa_- - \varkappa_+) & SC (\varkappa_- -
    \varkappa_+) & C^2 \chi_+ + S^2 \mu_-
  \end{array}\right) \label{eq:l2},
\end{equation}
upon introducing the notation
\begin{eqnarray}
  \varkappa_{\pm} & = & \Sigma_{\pm} \Delta_{\pm} \nonumber\\
  \mu_{\pm} & = & \Sigma^2_{\pm} + \Delta^2_{\pm} \nonumber\\
  \chi_{\pm} & = & \Sigma^2_+ \pm \Sigma^2_- \nonumber\\
  \kappa_{\pm} & = & \Delta_+^2 \pm \Delta^2_-, \label{eq:label} 
\end{eqnarray}
where
\begin{eqnarray}
  \Sigma_{\pm} & = & \frac{\sqrt{\lambda_{\pm 1}} + \sqrt{\lambda_{\pm 2}}}{2}
  \nonumber\\
  \Delta_{\pm} & = & \frac{\sqrt{\lambda_{\pm 1}} - \sqrt{\lambda_{\pm
  2}}}{2}, \label{eq:deltasigma} 
\end{eqnarray}
so that the constraints $\mu_- + \mu_+ = 1$ and $\kappa_+ + \varkappa_+ = 1$
are fulfilled.

\section{Choi matrices}

Given this matrix representation complete positivity can be checked using the
fact that the Choi matrices $\Phi^{1, 2}_{\tmop{Choi}} (t)$ associated to the
two maps are simply obtained by a suitable transposition of indexes
\begin{eqnarray}
  \Phi^{1, 2}_{\tmop{Choi}} (t) & = & \Lambda^{1, 2}_{ik, jl} (t) .
  \label{eq:mchoi} 
\end{eqnarray}
The expression of $\Phi^2_{\tmop{Choi}} (t)$ is given in
Eq.~(\ref{eq:Choiphi2}), and its positivity can be directly checked.

The matrix representation of $\Lambda^1_{ij, kl} (t)$ can be more simply
obtained exploiting the result Eq.~(\ref{eq:f2}), and therefore evaluating the
action of the Kraus operators $M_{\gamma \tmmathbf{\alpha}} (t)$ on the
elements $\{\Pi^1_S |i \rangle  \langle j| \Pi^1_S + \Pi^2_S |i \rangle
\langle j| \Pi^2_S \}_{i, j = 0, 1}$, leading to the result
\begin{equation}
  \Lambda^1 = \frac{1}{2} \left(\begin{array}{cccc}
    C^2 + 2 S^2 \mu_+ & 2 SC (\varkappa_- - \varkappa_+) & 2 SC (\varkappa_- -
    \varkappa_+) & C^2 + 2 S^2 \mu_-\\
    4 iS^2 \varkappa_+ & \begin{array}{c}
      C^2 - iCS (\mu_+ - \mu_-)
    \end{array} & - C^2 - iCS (\mu_+ - \mu_-) & 4 iS^2 \varkappa_-\\
    - 4 iS^2 \varkappa_+ & - C^2 + iCS (\mu_+ - \mu_-) & C^2 + iCS (\mu_+ -
    \mu_-) & - 4 iS^2 \varkappa_-\\
    C^2 + 2 S^2 \mu_+ & 2 SC (\varkappa_- - \varkappa_+) & 2 SC (\varkappa_- -
    \varkappa_+) & C^2 + 2 S^2 \mu_-
  \end{array}\right) \label{eq:l1},
\end{equation}
which further leads to the associated Choi matrix $\Phi^1_{\tmop{Choi}} (t)$
given in Eq.~(\ref{eq:Choiphi1}).

We note in particular that for the choice of system-environment state given by
Eq.~(\ref{eq:valori}) upon substituting in Eq.~(\ref{eq:label}) one recovers
the following Choi matrix Eq.~(\ref{eq:Choiphi2}) associated to $\Phi_2 (t)$
\begin{equation}
  \Phi_{\tmop{Choi}} = \frac{1}{2} \left(\begin{array}{cccc}
    \begin{array}{c}
      C^2 (1 + a_+ )\\
      + S^2 (1 + yb)
    \end{array} & SCb & - iS^2 b & \begin{array}{c}
      C^2 (1 + a_+)\\
      - iCS (yb - a_-)
    \end{array}\\
    SCb & \begin{array}{c}
      C^2 (1 - a_+)\\
      + S^2 (1 - yb)
    \end{array} & \begin{array}{c}
      - C^2 (1 - a_+)\\
      - iCS (yb + a_-)
    \end{array} & iS^2 b\\
    iS^2 b & \begin{array}{c}
      - C^2 (1 - a_+)\\
      + iCS (yb + a_-)
    \end{array} & \begin{array}{c}
      C^2 (1 - a_+)\\
      + S^2 (1 + yb)
    \end{array} & SCb\\
    \begin{array}{c}
      C^2 (1 + a_+)\\
      + iCS (yb - a_-)
    \end{array} & - iS^2 b & SCb & \begin{array}{c}
      C^2 (1 + a_+)\\
      + S^2 (1 - yb)
    \end{array}
  \end{array}\right) \label{eq:mjpa},
\end{equation}
where to simplify notation we have set
\begin{eqnarray}
  a_{\pm} & = & \frac{1}{2} \left( \sqrt{\frac{1 - (y + \chi)^2}{1 - y^2}} \pm
  \sqrt{\frac{1 - (y - \chi)^2}{1 - y^2}} \right) \nonumber\\
  b & = & \frac{\chi}{1 - y^2} . \nonumber
\end{eqnarray}

\section*{Acknowledgements}

B.V. acknowledges support from the EU Collaborative Project QuProCS (Grant
Agreement 641277) and by the Unimi TRANSITION GRANT - HORIZON 2020.

\end{document}